\documentclass[prl,twocolumn,showpacs]{revtex4}
\usepackage{graphicx}
\usepackage{amsmath}
\newcommand{\n}{\nonumber}
\newcommand{\bn}{\begin{eqnarray}}
\newcommand{\en}{\end{eqnarray}}
\newcommand{\eml}{\end{multline}}
\newcommand{\bml}{\begin{multline}}
\newcommand{\h}{\hspace}

\begin{document}

\title {What is Quantum in Quantum Pumping: The Role of Phase and Asymmetries}
 \author{Kunal K. Das$^1$, and  Tom\'{a}$\check{{\rm s}}$ Opatrn\'{y}$^2$}
 \affiliation{$^1$Department of Physics, Fordham University, Bronx, New York 10458, USA\\
 $^2$Optics Department, Faculty of Science, Palack\'{y} University, 17. Listopadu 50,
 77200 Olomouc, Czech Republic}

\date{\today }
\begin{abstract}
We show that quantum pumping does not always require a quantum
description or a quantum phase. Quantum pumping is shown to
encompass different types of processes, some of which intrinsically
rely on phase while others do not.  We also show that many pumping
processes have a hidden antisymmetric component that contributes
significantly to the instantaneous current at the terminals without
causing net charge transfer in a period. We have also computed the
exact pumped current for some cases over a full range of time
variation from adiabatic to non-adiabatic.

\end{abstract}
 \maketitle

As a viable candidate for generating directed biasless currents of
charge or spin through nanoscale devices, quantum pumping has been a
subject of numerous studies in recent years. The phenomenon has been
labeled by a multitude of characteristics: as a quantum interference
effect \cite{avron-math, buttiker-inelastic}, as a manifestation of
geometric phase \cite{zhou-mckenzie-geometric}, as a biasless
generator of dc current \cite{brouwer-1}, as a means of quantized
charge transfer \cite{thouless, qian-niu-1986} and as an
intrinsically quantum effect distinct from Coulomb blockade based
``classical pumps" \cite{aleiner-andreev,blaauboer-heller}. Equally
numerous have been the approaches used to study it, the
Landauer-Buttiker approach being the primary \cite{brouwer-1},
Keldysh Green's functions \cite{arrachea-green}, adiabatic
perturbation \cite{das-PRL,nanostructures}, Floquet theory
\cite{kim-floquet, buttiker-floquet}, and differential geometry
\cite{avron-math, avron-geometry-PRB,zhou-mckenzie-geometric}. An
unintended consequence of such varied labels and approaches has been
that a simple unambiguous way of understanding the essential
features of the phenomenon has been strangely elusive. Thus an
experiment reported as a demonstration of quantum pumping turned out
to be due to ac rectification \cite{switkes, brouwer-2};  other
experiments
\cite{shilton-SAW-1996,talyanskii-SAW-97,expt-fletcher-SAW} showing
features of quantum pumping \cite{entin-wohlman-SAW,
levinson-acoustoelectric} also accommodate alternate
\cite{talyanskii-SAW-97,robinson-SAW-classical}, even semiclassical,
explanations. A comprehensive view has also been hindered by the
difficulty of solving time-dependent problems, so that studies have
been confined to the adiabatic regime.

In this paper we strip down quantum pumping to its bare essentials,
and study it simply as a quantum mechanical process involving time
dependent potentials.  This has allowed us to identify several
essential features of the pumping mechanism that are often masked by
abstract formalisms, the details imposed by simulation of
experiments and the theoretical difficulty of going beyond the
adiabatic regime. For certain pumping scenarios, we will compute the
exact pumped current for arbitrary rates of time variation,
something that has not been possible in cases studied before. To
examine the adiabatic limit we use the expression \cite{das-PRL}
\bn J(t)\!= \!\frac{1}{2\pi}\!\int\!\! dx'\dot{V}(x'\!,t){\rm
Im}\{G^*(x'\!,x;E_F)
\partial_{x}G(x,x';E_F)\!\}\label{adiabatic-formula}\en
for the instantaneous adiabatic current. We set $\hbar=m=e=1$.
Although easily generalized this specific form assumes degenerate
temperatures typical of pumping experiments, reflected in its
dependence on the Fermi energy $E_F$. The description in terms of
the instantaneous Green's function $G(x,x';E_F)$ is due to
adiabaticity.

Since pumping is a direct consequence of time varying potential
acting in one-dimensional (1D) channels, we will examine the current
generated by representative time variations of a potential acting in
1D, where (i) the position, (ii) the strength or (iii) the shape of
the potential varies in time. In addition we will consider (iv) the
variation of a periodic lattice. We can capture the essential
features of all of these possibilities by using time-varying delta
function potentials.  Each case highlights certain distinct features
of quantum pumping, that we discuss, along with its quantum versus
classical nature.

\begin{figure}[b]\vspace{-7mm}
\includegraphics*[width=\columnwidth]{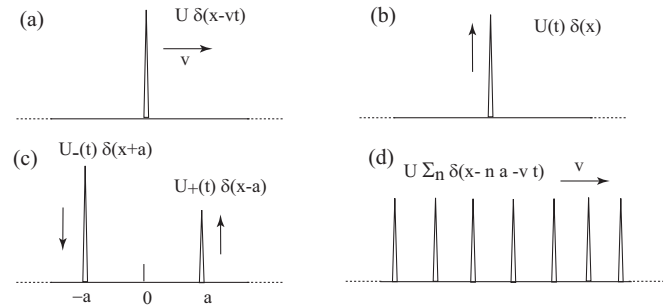}
\caption{The four typical time-varying potentials we consider (a) a
uniformly translating potential barrier, (b) a potential varying in
magnitude, (c) a potential changing its shape modeled by two
barriers varying in strength out of phase and (d) a uniformly
translating lattice, of spacing $a$.  For simplicity the elements of
the potentials are taken to be $\delta-$functions.} \label{Fig1}
\end{figure}

\begin{figure}[t]
\includegraphics*[width=\columnwidth]{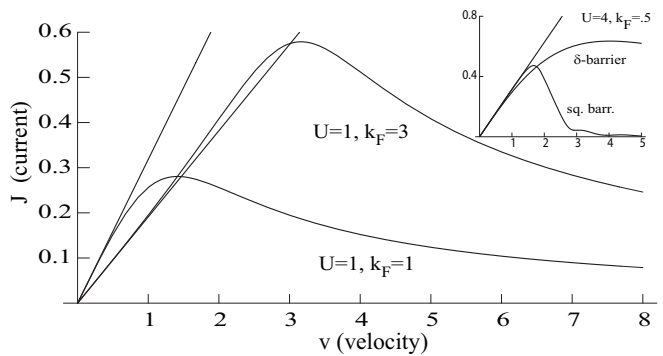}\vspace{-3mm}
\caption{The pumped current versus the velocity of a uniformly
moving $\delta$-potential barrier.  The curved lines are exact
solutions [Eq.~(\ref{1-delta})] and the straight lines are the
adiabatic solutions [Eq.~(\ref{1-delta-adiab})]. They agree for
$v\ll k_F$. The inset shows results for a moving square barrier
(height=width=2); the oscillations arise from quantum phase.}
\vspace{-5mm}\label{Fig2}
\end{figure}

\emph{Exact solution for a potential with uniform velocity:} We
start with a rare case where the pumped current can actually be
determined for arbitrary rate of change of the potential: a delta
potential moving with uniform velocity in a 1D channel,
$V(x,t)=U\delta(x-vt)$. At first sight this might not seem like a
quantum pumping process, but as will be seen, it has all the
features of pumping. A Galilean transformation yields the exact
stationary states
\bn e^{ivx-iv^2t/2-i(k-v)^2t/2}\left[e^{i(k-v)(x-vt)}
-\frac{e^{i(k-v)|(x-vt)|}}{1-i(k-v)/U}\right]\en
corresponding to the free-particle state $e^{ikx}$.  The
instantaneous current at any position is given by the $j(x,t)={\rm
Re}\{\psi^*_k(x,t)\frac{\hbar}{im}\partial_x\psi_k(x,t)\}$.
For right $e^{+ikx}$ and left $e^{-ikx}$ going waves, the
transmitted and reflected currents are : \bn j^\pm_T(k)=
\left[\frac{\pm k(k\mp v)^2}{(k\mp v)^2+U^2}\right],\h{.5cm}
j^\pm_R(k)=\frac{\mp(k\mp2v)U^2}{(k\mp v)^2+U^2}.\en
The net instantaneous current is obtained by averaging over the
current on both sides of the potential
\bn\label{1-delta} J=\h{-1mm}\int_{0}^{k_F}\h{-2mm} \frac{dk
}{2\pi}[j^+_T\!+\!j^-_R\!+\!
j^-_T\!+\!j^+_R]\!=\!\frac{U^2}{2\pi}\ln\h{-1mm}\left|\frac{U^2\!+\!(k_F\!+\!v)^2}
{U^2\!+\!(k_F\!-\!v)^2}\right|\en
and is independent of time. In the adiabatic limit $v\ll k_F$, a
Taylor expansion to linear order in $v/k_F$ yields
\bn\label{1-delta-adiab}
J&\simeq&\frac{1}{\pi}\frac{2k_FvU^2}{(k_F^2+U^2)}\label{ad-mov-pot}\en
which can be obtained identically using
Eq.~(\ref{adiabatic-formula}) or the Brouwer formula
\cite{brouwer-1} for the adiabatic pumped current, confirming that
this mechanism is an instance of quantum pumping by definition. It
is apparent that periodicity is not necessary for generating the
pumped current, but can be easily imposed by letting the potential
translate uniformly for a time $T$, switch off and repeat starting
at the initial point, much like a paddle in water. The strength of
the potential varying between on and off could be considered a
second parameter, but since neither the switching nor the
periodicity is essential for current generation, this is pumping
achieved by varying a single parameter, the velocity of the
potential. The current is plotted versus this parameter in
Fig.~\ref{Fig2}. In a finite system however, periodicity would be a
practical necessity to maintain a sustained current.  In the
adiabatic limit, Eq~(\ref{ad-mov-pot}), the pumped charge over a
time period $T$ is $J\times T$ which is independent of the velocity,
and depends only on the path traversed $X=v\times T$. This is
consistent with geometric interpretations of the adiabatic pumped
charge, however there is neither an energy gap, nor an essential
periodicity, so the pumped charge cannot be identified with
geometric phase in the Berry sense \cite{Berry}.

\emph{Classical versus Quantum}:  Having established that the
translating delta potential is an instance of quantum pumping, we
now demonstrate that it can be simulated classically as well without
invoking quantum mechanics, and that the phase of the wavefunction
plays no essential role. This latter point is important because,
quantum pumping has often been described as a quantum interference
effect, this shows that this is not always the case.

Consider a classical barrier moving with velocity $v$ in a one
dimensional channel, through a stream of particles which have a
uniform velocity distribution $\rho$ in the range $[-k_F,k_F]$.
Further assume that the moving barrier is a soft one possessing
velocity-dependent transmission and reflection probabilities $T(k)$
and $R(k)$.

A particle on the left of the barrier has the following possible
origins depending on its velocity $k$~: (i) if $k\in
[-k_F,-k_F+2v]$ it was transmitted from the right, (ii) if
$k\in[-k_F+2v,v]$ it was either transmitted from right \emph{or}
reflected at the barrier having come from the left and (iii) if
$k\in[v,k_F]$, it never interacted with the barrier, since its
moving faster than the barrier and is still behind it. The net
current on the left of the barrier is
{\small \bn J_L=\rho\h{-1mm}\int_{-k_F}^{-k_F+2v}\h{-11mm}dk
T(k)k+\rho\h{-1mm}\int_{-k_F+2v}^{v}\h{-10mm}dk [T(k)+R(-k+2v)]k
+\rho\h{-1mm}\int_{v}^{k_F}\h{-5mm}dk k.\en}
A similar expression can be obtained for the current $J_R$ on the
right of the barrier, based upon an analogous argument, with the one
difference that the full range for $k$ is $[-k_F,k_F+2v]$ because of
the particles that have been reflected with a boost in the velocity
due to the barrier.

If the scattering probabilities are chosen to coincide with the
quantum mechanical probabilities and $\rho=1/2\pi$, then $J_L+J_R$
gives the same current as in Eq.~(\ref{1-delta}). This demonstrates
that ironically quantum mechanics is not always essential for
quantum pumping. One can simulate this situation with classical
physics, with a barrier that transmits particles or reflects with
specific momentum dependent probabilities. The essentially quantum
ingredient, the phase, plays no physical role, inasmuch as the
velocity change for the reflected particles can be understood simply
from conservation of momentum in classical mechanics.
A finite size potential, say a square barrier instead of a delta
function, displays some effects of phase, where the phase shift
incurred in transmitting through the potential, creates current
oscillations (Fig.~\ref{Fig2} inset). However, the arguments above
are still applicable with appropriate scattering probabilities,
therefore the effect can still be simulated classically.

\emph{Potential varying in strength}: Other important, but little
known, features of quantum pumping can be highlighted by considering
a different kind of time variation, shown in Fig.~1(b), where the
potential varies in strength, for which
Eq.~(\ref{adiabatic-formula}) gives the adiabatic pumped current
\bn V(x)=U(t)\delta(x);\h{5mm}J(x)={\rm sign(x)}\frac{1}{2
\pi}\dot{U}\frac{k}{k^2+U^2}\label{growing-current}\en
for $x$ far from the potential. Over a full time period this would
give zero net current, being an exact differential. Yet a current
detector at either terminal in a nanoscale circuit would measure a
continuous non-zero current. Current alternately flows out into the
leads, as the potential grows, and flows in as it diminishes. The
outflow and the inflow cancel over a closed cycle. This makes it
clear that an averaging over all terminals is crucial in determining
a genuine pumped current.

Contrasting this with the previous case of uniformly translating
potential highlights another essential point; in a two terminal
circuit the current can have contributions that are symmetric or
antisymmetric with respect to the terminal, or the side of the
potential, where the current is measured. For the translating
potential the adiabatic current was completely symmetric and there
was a net flow, while for time-varying magnitude, it is completely
antisymmetric and hence there is no net flow.

Finally we note that a potential varying in strength effectively
serves as a source or a sink for charge in the leads, which can also
be understood classically: a barrier emerging from the bottom of a
channel containing fluid would cause outward flow due to
displacement.

\begin{figure}[b]
\includegraphics*[width=\columnwidth]{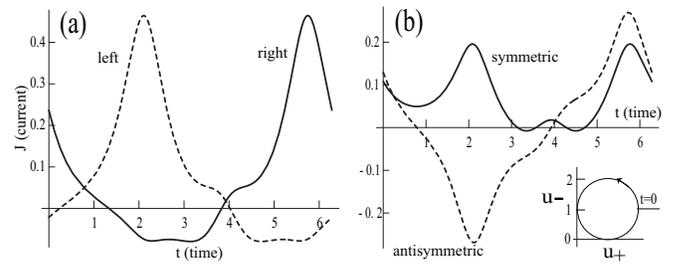}
\caption{The instantaneous current $J(t)$ as a function of time over
a period $t\in(0,2\pi)$, for $k_F=1, a=1$. The inset shows the time
variation of pumping parameters $U_-=1+\cos(t)$ and $U_+=1+\sin(t)$.
Plot (a) shows the current flowing into the left terminal $(x<0)$
and the right terminal $(x>0)$ and plot (b) shows the symmetric and
antisymmetric part of the current. It is apparent that the
antisymmetric part would average to zero over a full cycle.}
\label{Fig3}
\end{figure}

\emph{Potential changing its shape}: A quantum pump that operates by
shape variation of a potential can be examined by considering the
commonly used turnstile model \cite{turnstile}, where two potentials
separated by a finite space (2a), vary in strength out of phase with
each other $V(t)=\sum_{\pm}U_{\pm}(t)\delta(x'\mp a)$. For an
observation point $|x|\rightarrow \infty$ far from the potential,
the adiabatic current is
\bn J(x,t)&=& \frac{k}{2\pi}\ {\rm sgn}(x)\sum_{x'=\pm
a}\dot{U}_{\pm} |G(x,x')|^2 \n\\
ikG(x,x')&=&e^{ik|x-x'|}+e^{ik|x-a|} \Delta_+(x')+
e^{ik|x+a|}\Delta_-(x')\n\\
\Delta_{\pm}(x')&=&\frac{r_\pm}{D}\left(e^{ik|x'\mp a|}+ r_\mp
e^{ik|x'\pm a|+2ika}\right)\en
with Green's function $G(x,x')$, and $D=1-r_+r_- e^{i4ak}$ where
$r_{\pm}(k)=U_{\pm}/(U_{\pm}^2+k^2)$ are the reflection
probabilities at the two delta potentials. The current at the right
($J_R$) and the left ($J_L$) terminals can be separated into
(i)$J_S=\frac{1}{2}(J_R+J_L)$ symmetric and
(ii)$J_A=\frac{1}{2}(J_R-J_L)$ antisymmetric, with respect to $x$
\bn J_S(t)&=&\frac{1}{\pi k}\sum_{x'=\pm
a}\dot{U}_{\pm}\left[\sin(2ka){\rm
Im}\{\Delta_+^*\Delta_-\}\right.\\&+&\left.\sin[k(x'-a)]{\rm
Im}\{\Delta_+\}+\sin[k(x'+a)]{\rm Im}\{\Delta_-\}\right]\n\\
J_A(t)&=&\frac{1}{\pi k}\h{-2mm}\sum_{x'=\pm
a}\h{-2mm}\dot{U}_{\pm}\left[{\textstyle \frac{1}{2}(1+
|\Delta|^2)}+\cos(2ka){\rm
Im}\{\Delta_+^*\Delta_-\}\n\right.\\&+&\left.
\cos[k(x'-a)]{\rm Im}\{\Delta_+\}+\cos[k(x'+a)]{\rm
Im}\{\Delta_-\}\right]\n\en
with $|\Delta|^2=|\Delta_+|^2+|\Delta_-|^2$.   The current at the
left and right terminals are plotted in Fig.~\ref{Fig3}(a) for a
full period of cyclic variation of $U_{\pm}(t)$, and the symmetric
and antisymmetric parts are plotted in Fig.~\ref{Fig3}(b).  As
discussed above only the symmetric part leads to a net current over
a cycle, the plot Fig.~\ref{Fig3}(b) shows that a time integral over
the antisymmetric part vanishes. Application of Green's theorem
$\oint dt[J_+ \dot{U}_+-J_+ \dot{U}_-]\rightarrow \int
dU_+dU_-[\frac{\partial J_-}{\partial U_+}-\frac{\partial
J_+}{\partial U_-}]$ for a full cycle of $U_{\pm}(t)$ to each of the
currents $J_R$, $J_L$, $J_S$ and $J_A$ separately shows that the
\emph{integrand} $[\frac{\partial J_-}{\partial U_+}-\frac{\partial
J_+}{\partial U_-}]$ of the corresponding surface integral vanishes
for $J_A$, but has the same value for $J_R$, $J_L$ and $J_S$.

The fact that it is the integrand itself which vanishes, and not
just the integral, is significant; because in the surface integral
form, as for instance in the Brouwer formula \cite{brouwer-1}, the
contribution of the antisymmetric component is completely lost while
the current at the terminals is exactly synonymous with the
symmetric part. Yet, as Fig.~\ref{Fig3} shows, the instantaneous
current at each terminal can often be dominated by the antisymmetric
part during the course of a cycle.

These plots  reveal another fact, that the pumped current is not a
dc current really, but more like an ac current with an offset
resulting in a net transfer of charge over a period. Even the
symmetric part that contributes to the net current can reverse
direction over the cycle.

Much of this mechanism can also be understood classically. As we
already established, current would be generated even if only one of
the potentials is changing. The presence of the other barrier
hinders flow on one side, leading to an imbalance in current flowing
into the two terminals. The asymmetry of the variation of the two
potentials causes that imbalance to add up to a net transport over a
cycle. If the potentials varied in sync, that imbalance would cancel
itself out over a full cycle.

Unlike the previous two mechanisms, phase plays a significant role
here, as can be concluded from the presence of the $\sin(2ka)$
factor in the non-vanishing symmetric part of the current.  The
current would vanish identically if the electron wavevector
satisfies $2ka=n\pi$, which  physically means that as each potential
changes, the electron propagation is not be affected by the other
potential if it is at a nodal distance for the wavefunction. Varying
the spatial separation of the potentials should lead to an
observable sinusoidal variation of the net current, but this  should
not be confused with the sinusoidal variation arising from the
temporal phase lag among the time varying parameters discussed
elsewhere \cite{brouwer-1}. In any case, the tangible effect of the
phase of the wavefunction makes this pumping mechanism a manifestly
quantum effect.

\emph{A moving lattice potential}: Spatially periodic potentials
have been used in the context of quantum pumping since the initial
paper by Thouless \cite{thouless} and also recently
\cite{altchuler-science}; for filled bands charge transfer was shown
to be quantized . We will now examine if that is necessarily a
quantum effect, by studying a uniformly moving evenly spaced array
of delta potentials, effectively a moving Kronig-Penney (KP)
potential [Fig.~\ref{Fig1}(d)], of period a, system size $L$. This
can also be exactly solved for all rates of time variation, by using
a Galilean transformation. For a stationary state $\alpha e^{ik
x}+\beta e^{-ik x}$ satisfying the KP boundary conditions, the
current for each $k$ is:
\bn
j(k,t)=\frac{v}{L}+2v\alpha\beta\sum_{n}\left[\cos(2k\{x-na+(n-1)vt\})\h{2mm}\right.\\
\left.\times
\theta(x-n_-(a-vt))\theta(n_+(a-vt)-x)\right]+(\alpha^2-\beta^2)k
\n\en
with $n_{\pm}=n\pm\frac{1}{2}$. The net current is given by
integrating over allowed vectors $\int_{-k_F}^{k_F}
\frac{dk}{2\pi}j(k,t)$, but the essential conclusions can be reached
without doing the integral. Instead we do a time integral over a
period $[0,a/v]$ to get the charge transfer per period for each
wavevector $k$
\bn q(k,T)=\int_{0}^{a/v}
j(k,t)=\frac{a}{L}+(\alpha^2-\beta^2)\frac{ka}{v}\label{kp-charge}\en
In the case of \emph{filled} bands, summation over the states would
lead to a vanishing of the last term, while the surviving first term
gives the fraction of the particles in the interval $a$, which would
be an integer, giving quantized charge transport.  The second term
in Eq.~(\ref{kp-charge}) contributes for partially filled bands, and
is proportionate to the time $T=a/v$, the duration of the flow.

It is clear that the charge quantization for filled bands is
equivalent to the trivial result one would get classically if one
moved a lattice containing uniformly distributed classical
particles; transport by one spatial period would lead to transfer of
the number of particles in each site. The second term is however due
to quantum tunnelling and the $cosine$ dependence of the current is
an effect of quantum phase, which however does not contribute to the
net charge transfer over a period commensurate with the spatial
period.  These results are exact and not a consequence of an
adiabatic approximation.

\emph{Conclusions}: From our discussions above we conclude that
quantum pumping actually encompasses a variety of distinct
processes, described by the same quantum mechanical expressions.
However not all are strictly quantum effects, some can be simulated
by classical mechanisms while others can be explained only in a
quantum picture;  the role of the phase of the wavefunction is the
crucial differentiator. This diversity of pumping processes explains
some of the conflicting pictures of experimental results involving
surface acoustic waves
\cite{shilton-SAW-1996,talyanskii-SAW-97,expt-fletcher-SAW}, where
both classical methods (a dragging mechanism)
\cite{talyanskii-SAW-97,robinson-SAW-classical} and quantum picture
(a quantum pump) \cite{entin-wohlman-SAW, levinson-acoustoelectric}
could describe experimental results. Also contrary to statements
often used in the literature, quantum pumping is not always a
quantum interference effect, as evidenced by the fact that phase
does even have to play a role. Even temporal periodicity is not
necessary to generate a current, but is more of a practical
necessity to maintain a sustained current in a finite system.

We also showed that the pumped current in a two terminal setup has
symmetric and antisymmetric components. While only the symmetric
component may cause net charge transfer and is  manifest in most
commonly used formalisms, the hidden antisymmetric component
contributes significantly to the instantaneous current at each
terminal. Furthermore, the pumped current is often like an ac
current with an offset that can reverse direction over a cycle
rather than a dc current as it is often labeled.  We also presented
results for certain quantum pumping mechanisms over all regimes of
time variation instead of just the adiabatic regime commonly
studied, clearly showing the nature of the transition from the
adiabatic to non-adiabatic regimes.

We gratefully acknowledge the support of the Research Corporation,
and of GA\v CR (GA202/05/0486) and M\v SMT (MSM~6198959213).

\end{document}